\documentclass[apsspec,showpacs,twocolumn,groupedaddress,nofootinbib,lengthcheck,preprintnumbers]{revtex4}
\usepackage{amsmath}
\usepackage{amssymb}
\usepackage{graphics}
\usepackage[dvips]{graphicx}
\usepackage{graphicx}
\usepackage[usenames]{color}

\def\simge{\mathrel{%
       \rlap{\raise 0.511ex \hbox{$>$}}{\lower 0.511ex \hbox{$\sim$}}}}
\def\simle{\mathrel{
       \rlap{\raise 0.511ex \hbox{$<$}}{\lower 0.511ex \hbox{$\sim$}}}}

\newcommand \beq{\begin{eqnarray}}
\newcommand \eeq{\end{eqnarray}}

\usepackage[colorlinks=true,linktocpage=true,linkcolor=blue,citecolor=blue]{hyperref}

\begin{document}
\title{Inverse Tritium Beta Decay with Relic Neutrinos, Solar Neutrinos, and a
$^{51}$Cr Source}
\author{Jen-Chieh Peng and Gordon Baym}
\affiliation{\mbox{Illinois Center for Advanced Studies of the Universe}\\}
\affiliation{\mbox{Department of Physics, University of Illinois, 1110
  W. Green Street, Urbana, IL 61801} \\
}

\date{\today}

\begin{abstract}
The  inverse tritium beta decay (ITBD) reaction, $\nu_e + ^3$H $\to e^- + ^3$He, is a promising 
experimental tool for observing relic neutrinos created in the 
early Universe. 
This reaction has been selected
by the PTOLEMY experiment for the search of relic neutrinos.
Despite its potential,
the ITBD reaction induced by any sources of neutrinos has yet to
be observed. We show that an intense $^{51}$Cr radioactive neutrino source
is suitable for observing the ITBD reaction for the
first time.  As the Sun is another source of intense electron neutrinos, we also
examine the ITBD reaction rate from solar neutrinos.
Based on our recent studies on the evolution of the helicity of relic neutrinos,
we further present the ITBD rate for capturing relic neutrinos
as a function of neutrino mass hierarchy, the Dirac versus Majorana nature
of neutrino, and the mass of the lightest neutrino. 

\end{abstract}

\maketitle

\section{Introduction}

The standard cosmological model predicts that relic neutrinos decouple
 at about one second after the Big Bang~\cite{Zeldovich,Dolgov},
much earlier than the decoupling time, $\sim 3.8 \times 10^5$ years, of 
the Cosmic Microwave Background (CMB). While the detection of the 
CMB has revolutionized cosmology, eventual observation of the predicted Cosmic 
Neutrino Background (C$\nu$B) would provide valuable snapshots of the Universe at
an epoch much earlier than that of the CMB.

The most promising method for detecting the C$\nu$B, proposed by Weinberg
in 1962~\cite{Weinberg}, is capture of
electron neutrinos on a radioactive target, e.g., the 
inverse tritium beta decay (ITBD) reaction, 
$\nu_e + ^3$H $\to e^- + ^3$He.  Efforts to detect relic neutrinos via the ITBD reaction are
ongoing at the PTOLEMY experiment~\cite{Ptolemy}.  

  But not only has the C$\nu$B not been detected, remarkably, the 
ITBD reaction itself
has never been observed.    Detection of ITBD from characterized neutrino sources, both local and solar, 
would, aside from validating the theory of the ITBD reaction,  
test the performance of detector systems for relic neutrinos.  
In this paper we propose using an intense $^{51}$Cr 
source of electron neutrinos, from the electron capture reaction 
$e^- + ^{51}$Cr$ \to ^{51}$V$+ \nu_e$, to observe the ITBD reaction.  As the Sun is an intense source of electron neutrinos, we 
examine as well the
expected ITBD reaction rate initiated by solar neutrinos. 
Solar neutrinos could potentially set an irreducible
background for C$\nu$B detection in the ITBD reaction analogous to 
the ``neutrino floor" imposed by solar neutrinos for direct dark matter 
searches~\cite{vergados}. 
Comparing the ITBD rates from a $^{51}$Cr source and solar neutrinos with 
the expected rates originating from the C$\nu$B, we find that while the background for the C$\nu$B detection 
from solar neutrinos is negligible, the $^{51}$Cr source could yield the first observation of the ITBD reaction.

  The energy release, $Q$, of the ITBD reaction is positive, allowing
relic neutrinos with practically zero energy to be detected.
Capture of relic neutrinos by ITBD 
has the distinct signature of mono-energetic electrons
separated
from the endpoint of the tritium beta decay spectrum by twice the neutrino mass, $m_\nu$.
As discussed below, the ITBD cross section depends sensitively
on the helicity of the incident neutrino. We recently examined the 
evolution of the helicity of relic
neutrino in cosmic and galactic magnetic fields as well as in cosmic
gravitational inhomogeneities, and found a physically significant
probability of helicity reversal~\cite{Baym1,Baym2}.  
We further discuss here the consequences of this helicity evolution
of relic neutrinos on the rate of their detection via ITBD.

The predicted C$\nu$B density, summed over all flavors
of active neutrinos and antineutrinos,  is $\sim$ 338/cm$^3$, comparable to the CMB density
$\sim$ 411/cm$^3$~\cite{Dolgov}. The present 
temperature of the C$\nu$B, related to the
temperature of the CMB by $T_{\textrm{C$\nu$B}} = 
(4/11)^{1/3} T_{\textrm{CMB}}$, is 1.945 K  
corresponding to an energy of $1.676 \times 10^{-4}$ eV. Given the current mass-squared differences, $\Delta m_{ij}^2$, between neutrino mass states
$i$ and $j$: 
$\Delta m_{21}^2 = (7.50^{0.19}_{0.17}) \times 10^{-5}$ eV$^2$ and
$|\Delta m_{32}^2| = (2.52 \pm 0.04) \times 10^{-3}$ eV$^2$~\cite{RPP},
 at least two of the three neutrino mass eigenstates in the
C$\nu$B are presently
non-relativistic. Detection of the relic neutrinos poses the special 
challenge that, regardless of their origin, non-relativistic 
neutrinos have never been observed. 

\section{Inverse Tritium Beta Decay from $^{51}$Cr Source and Solar Neutrinos}

\begin{figure}
\includegraphics*[width=1.1\linewidth]{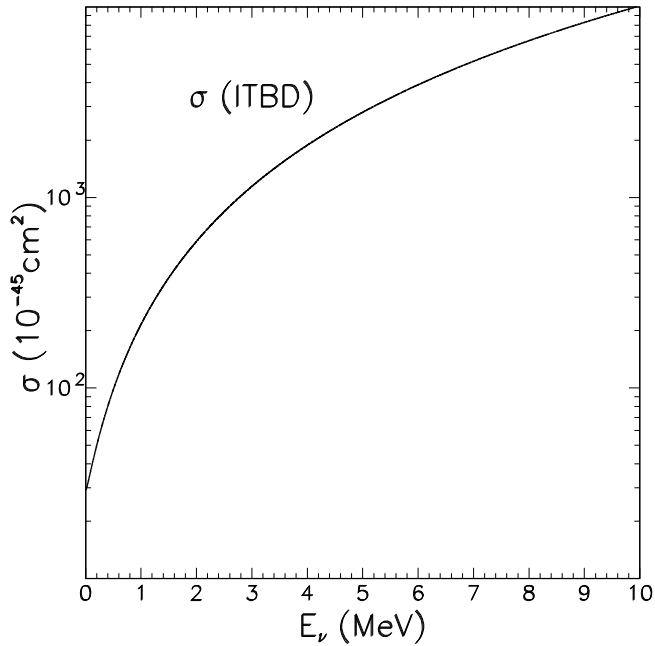}
\caption{
The ITBD cross section, Eq.~(\ref{Cr1}), versus incident electron
neutrino energy $E_\nu$, for massless neutrinos.   
}
\label{itbd_sigma}
\end{figure}

The cross section for capturing an electron neutrino with energy $E_\nu$ and
helicity $h$ on a tritium atom is  \cite{long}
\beq
\sigma^h (E_\nu) &=& \frac {G_F^2}{2\pi v} |V_{ud}|^2 F(Z,E_e)
\frac{m_{_{^3 \rm He}}}{m_{_{^3\rm H}}}
E_e p_e \nonumber \\
&&\hspace{20pt} \times (\langle f_F \rangle^2 + (g_A/g_V)^2 \langle g_{GT} \rangle^2) A^h,
\label{Cr1}
\eeq
where $v$ is the neutrino velocity; $V_{ud}$ 
is the up-down quark element of the CKM matrix and $F(Z,E_e)$ is the Fermi 
Coulomb correction for the $e - ^3$He system
\beq 
F(Z,E_e)=\frac{2\pi \eta}{1-e^{-2\pi\eta}};~~~\eta=Z\alpha E_e/p_e,
\label{Cr1a}
\eeq
where $Z=2$ is the atomic number of $^3$He and $\alpha$ is
the electromagnetic fine structure constant.  The nuclear form factors 
for Fermi and
Gamow-Teller transitions are
\beq
\langle f_F \rangle^2 \approx 0.9987,~~~\langle g_{GT} \rangle^2 \approx 2.788,
\label{Cr1b}
\eeq
with $g_V=1$ and $g_A =1.2695$. 
The neutrino helicity-dependent factor is 
\beq
A^{\pm} = 1\mp \beta,
\label{Cr2}
\eeq 
where $\beta = v/c$.
For relativistic neutrinos ($\beta \to 1$),
the ITBD reaction is dominantly from neutrinos with negative helicity. 
For slowly-moving
relic neutrinos ($\beta \to 0$), Eq.~(\ref{Cr2}) implies that the ITBD
cross section is nearly identical for the two neutrino helicity states.
Figure~\ref{itbd_sigma} displays the ITBD cross section versus the total electron-neutrino
energy $E_\nu$, with the neutrino mass neglected, so that
$\beta =1$, $A^+ = 0$ and $A^- = 2$. 
The pronounced energy dependence of the ITBD cross section 
in Fig.~\ref{itbd_sigma} is largely from the $E_e p_e$ factor 
with a minor contribution from the Fermi Coulomb correction.

\begin{figure}
\includegraphics*[width=1.1\linewidth]{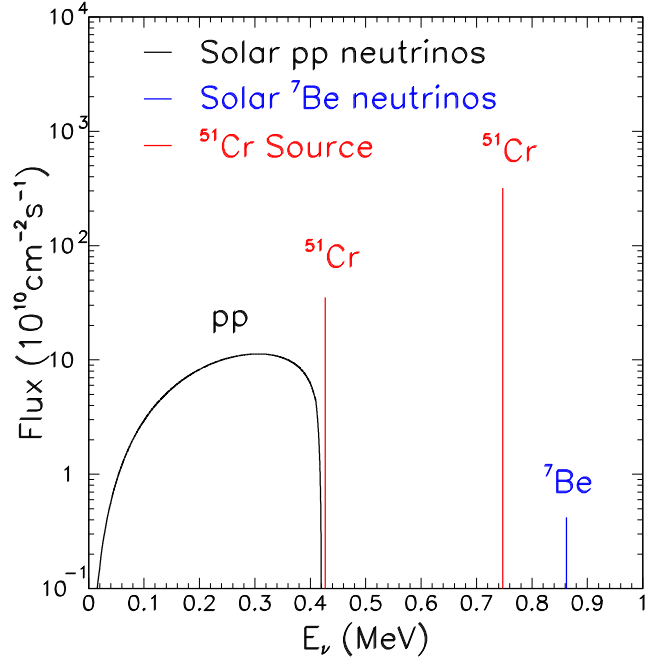}
\caption{
Flux of various sources of neutrinos versus neutrino energy.   The 
four $^{51}$Cr lines are shown in two groups, 0.427 + 0.432 MeV and  
0.747 + 0.752 MeV.
For $^{51}$Cr neutrinos and $^7$Be solar neutrinos, the fluxes are in unit
of $10^{10}/\mbox{cm}^2/\mbox{s}$.
For the $pp$ solar neutrinos, the flux is expressed
in units of $10^{10}\mbox{cm}^{-2}\mbox{s}^{-1}\mbox{MeV}^{-1}$ to account for 
the continuous energy spectrum.  
}
\label{itbd_flux}
\end{figure}

We consider two
possible electron neutrino sources besides relic neutrinos
for initiating the ITBD reaction,  $^{51}$Cr and the Sun.  
Intense $^{51}$Cr sources
have been utilized by several neutrino experiments, including
GALLEX~\cite{gallex}, SAGE~\cite{sage}, and BEST~\cite{best}. 
In particular, a search for a sterile neutrino was recently reported by the BEST
Collaboration using a 3.4-MCi $^{51}$Cr source~\cite{best}.
Another intense source of electron neutrinos is the Sun. 
Below, we first examine the
expected ITBD rate utilizing an intense $^{51}$Cr source, and then calculate
the ITBD rate from solar neutrinos.   It is important
to evaluate the expected ITBD rate to understand the inevitable ``background" 
events initiated by solar neutrinos.

The energy spectrum of electron neutrinos
from a $^{51}$Cr source consists of four discrete lines with
energies 0.427, 0.432, 0.747, and 0.752 MeV with respective branching ratios of
9.0, 0.9, 81.6, and 8.5 percent. 
Figure~\ref{itbd_flux} shows the 
average neutrino flux at the target region from the two groups of neutrinos
at 0.427 and 0.432 MeV and 0.747 and 0.752 MeV, from a 3-MCi $^{51}$Cr 
source is placed at an average distance
of 50 cm from the tritium target.

As a specific example of a feasible setup, a compact 3.4-MCi $^{51}$Cr 
source as fabricated for the BEST
experiment~\cite{best} could be placed at a 
distance of 50 cm, or less, from the tritium target. This source
consists of 26 enriched (97\%) metallic $^{50}$Cr
disks, each with a diameter of 8.4 or 8.8 cm and a thickness of 0.4 cm.
After irradiation at the SM-3 reactor at the Research Institute of
Atomic Research with thermal neutrons for $\sim 100$ days,
the disks were placed in a hermetic stainless-steel capsure covered with a
shield of 3-cm thick tungsten alloy. The entire assembly of the $^{51}$Cr
source is 16 cm in diameter and 22.6 cm in height. 

\begin{figure}
\includegraphics*[width=1.1\linewidth]{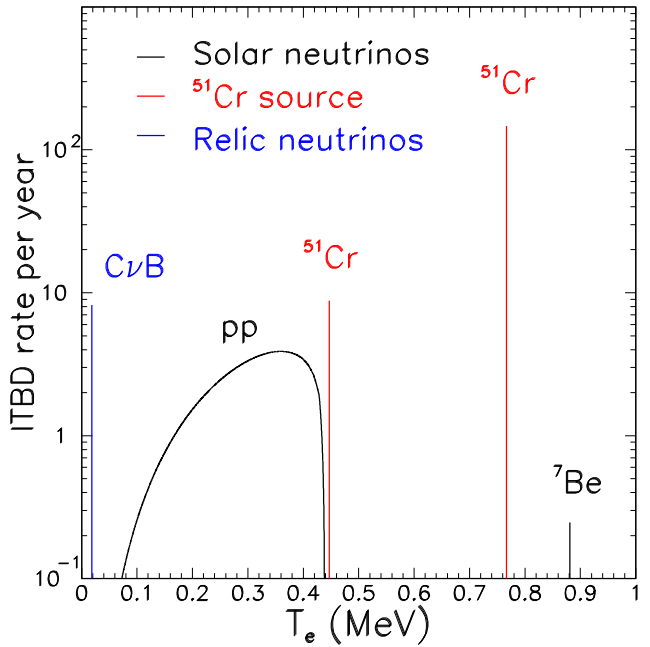}
\caption{
Rates of ITBD per year with a 100 g tritium target for various sources of 
neutrinos versus the kinetic energy of electron emitted in the ITBD reaction.
The $pp$ solar neutrino rate is in units of (Year-MeV)$^{-1}$.
The relic neutrinos are assumed to be of Majorana type 
for the rate calculation.
}
\label{itbd_rate}
\end{figure}

We evaluate the ITBD rates on a 100 g tritium target, 
as proposed by the PTOLEMY 
collaboration~\cite{Ptolemy}. The total ITBD rate per year for 
a 3-MCi  $^{51}$Cr source located 50 cm from the tritium target
is shown in Fig.~\ref{itbd_rate}, plotted as a function of kinetic energy of the electron emitted in 
the ITBD reaction. The ITBD rates from various sources
of electron neutrinos are listed in Table~\ref{itbd_tab}.

\begin{table}[tbp]
\caption {ITBD rate for various sources of electron neutrinos, together with the electron
kinetic energies, $T_e$.   The relic neutrinos
are assumed to be Majorana in the rate calculation.}
\begin{center}
\begin{tabular}{|c|c|c|c|}
\hline
\hline
Source & $T_e$ (MeV) & Rate (1/year) \\
\hline
$^{51}$Cr 0.427~+~0.432 MeV $\nu_e$ & 0.447 & 8.8 \\
$^{51}$Cr 0.747~+~0.752 MeV $\nu_e$ & 0.767 & 147.0 \\
Solar $pp$ $\nu_e$ & 0.0186 to 0.44 & 0.8 \\
Solar $^7\mbox{Be}$ $\nu_e$ & 0.881 & 0.23 \\
Relic  $\nu_e/\bar \nu_e$ & 0.018 & 8.2 \\
\hline
\hline
\end{tabular}
\end{center}
\label{itbd_tab}
\end{table}

    We turn to solar neutrinos, considering the most dominant
components originating from the $p+p \to d + e^+ + \nu_e$ 
reaction ($pp$ neutrinos) and the $^7$Be neutrinos at 0.862 MeV
from the $e^- + ^7$Be $\to \nu_e +^7$Li reaction.
The fluxes of $pp$ and $^7$Be solar neutrinos~\cite{bahcall} are
shown in Fig.~\ref{itbd_flux}.
The flux reduction of $pp$ solar 
neutrinos by a factor of 0.55 from neutrino oscillations, as measured by
Borexino~\cite{Borexino1}, is taken into account.  Similarly, we include 
a reduction
factor 0.53 from neutrino oscillations for the $^7$Be neutrinos.
As the $pp$ neutrinos have a continuous neutrino energy distribution from 0 to
0.4 MeV, we give the $pp$ solar neutrino flux in units of
$10^{10}\mbox{cm}^{-2}\mbox{s}^{-1}\mbox{MeV}^{-1}$. 

  Figure~\ref{itbd_flux} indicates that
the flux of the 0.747 + 0.752 MeV neutrinos from the $^{51}$Cr 
source is more than
an order of magnitude higher than the integrated flux of solar
$pp$ neutrinos. Together with the higher ITBD cross sections for
higher energy neutrinos from the $^{51}$Cr source,
one expects a significantly larger rate for the ITBD process from the
$^{51}$Cr source than from solar neutrinos.
Figure~\ref{itbd_rate} shows the total ITBD rate per year
with a 100 g tritium target, as proposed by the PTOLEMY
Collaboration, for neutrinos from $^{51}$Cr and the solar
$pp$ reaction. The ITBD rate in Fig.~\ref{itbd_rate} is
plotted as a function of the kinetic energy of the electron emitted in the ITBD
reaction. 

\section{Inverse Tritium Beta Decay from Relic Neutrinos}

The capture rate for a relic neutrino in mass eigenstate $i$ with 
energy $E_\nu$ and helicity $h$ on a tritium target of
mass $M_t$ is
\beq
\frac {d\Gamma^h_i (E_\nu)} {dE_\nu}  = |U_{ei}|^2 \frac{d\rho (E_\nu)}
{dE_\nu} v_i 
\sigma^h_i (E_\nu) N_t,
\label{Relic1}
\eeq
where $U_{ei}$ is the mixing matrix element between neutrino
mass eigenstate $i$ and $\nu_e$; $\rho(E_\nu)$ is the relic neutrino
density distribution
\beq
\frac{d^3\rho(E_\nu)}{d^3 p_\nu} \propto 
\frac{1}{e^{p_\nu/T_{\textrm{C$\nu$B}}}+1},
\label{Relic2}
\eeq
where $p_\nu$ is the relic neutrino momentum; $v_i$ is the velocity of
relic neutrino and $\sigma^h_i(E_\nu)$ is the ITBD cross section given
in Eq.~(\ref{Cr1}).  The number of tritium nuclei in the target is
$N_t = M_t/m_{_{^3\textrm{H}}}$.

The helicity dependence of the ITBD cross section
makes it crucial to take the helicities
of relic neutrinos into account in predicting the ITBD rate.  We recently investigated
the evolution of the helicities of relic neutrinos
as they propagate through cosmic magnetic fields and gravitational
inhomogeneities~\cite{Baym1,Baym2}, and
briefly summarize our findings.

In the early Universe, neutrinos were in chiral eigenstates. When
decoupled at a temperature of $\sim 1$ MeV, they were highly relativistic
and the chiral eigenstates essentially coincide with helicity eigenstates.
Neutrinos decouple effectively in negative helicity states, while
antineutrinos decouple with positive helicity.
On their long journey to
Earth, their helicities can be modified by two
effects. First, while their momentum and spin vectors are both bent by gravitational
forces acting transverse to their direction of motion,
the rotation angle of the spin vector of a non-zero mass neutrino 
is less 
than that of the momentum vector ~\cite{Baym2,Oleg}.
Thus a neutrino of negative helicity
at the time of decoupling would acquire an amplitude to have a positive helicity
component as it propagates through the Universe. Similarly, a positive helicity
antineutrino at decoupling would accumulate a non-zero negative-helicity
amplitude.   Reference~~\cite{Baym2} calculates in detail the 
root-mean-square bending 
angles of relic neutrino momentum 
$\langle (\Delta \theta_p)^2 \rangle^{1/2}$,
spin $\langle (\Delta \theta_S)^2 \rangle^{1/2}$
and spin relative to momentum $\langle (\Delta \theta)^2 \rangle^{1/2}$, 
with the gravitational inhomogeneities of the Universe measured
in the Planck experiment as input~\cite{Planck1},

  Another source for modifying relic neutrino helicities are the cosmic
and galactic magnetic fields. While magnetic fields do not affect the neutrino momentum, the spin of a neutrino with a nonzero diagonal magnetic
moment, as non-zero mass Dirac neutrinos should have, precesses in magnetic fields, hence modifying the neutrino helicity.  Both Majorana and Dirac neutrinos can have transitional magnetic moments, but only Dirac neutrinos
can have a diagonal magnetic moment.\footnote{The reported observation 
(now withdrawn~\cite{XENONnT}) of an excess of low-energy electron events by
XENON1T~\cite{XENON1T} prompted the suggestion that solar neutrinos
could have a large magnetic moment, of order
$\sim 1.4 - 2.9 \times 10^{-11} \mu_B$~\cite{Miranda, Babu}, which is
compatible with the upper limit of $\mu_{\nu_e} < 2.8 \times 10^{-11} \mu_B$
set by the Borexino experiment~\cite{Borexino}.}

As relic neutrinos approach the Earth, they encounter the magnetic fields
of the Milky Way, with magnetic field $B_g \sim 10 \mu$G. As the galactic magnetic
fields change orientation over a coherence length, $\Lambda_g$,
of order kpc, the spin
orientation of relic neutrino undergoes a random walk through the galaxy.
As shown in Refs.~\cite{Baym1,Baym2}, while the gravitational 
bending of a neutrino
spin with respect to its momentum is well below that produced by a magnetic
moment of the magnitude of the Borexino upper limit~\cite{Borexino},  
the cumulative
rotation of a Dirac relic neutrino from the cosmic magnetic
fields is  comparable to that from the Milky Way ~\cite{Baym1}.

\begin{figure}[t]
\centering
\includegraphics*[width=1.0\linewidth]{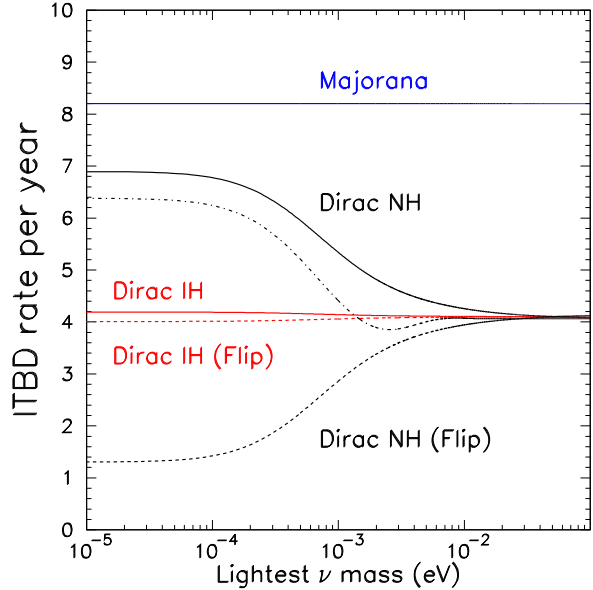}
\caption{
Rates of ITBD per year for relic neutrinos on a 100 g tritium target.
The rates versus mass of the lightest neutrino are shown as solid curves
for the normal (NH) and inverted (IH) mass hierarchies for Dirac and
Majorana neutrinos. The dashed curves show
the case of complete helicity flip from left to right handed neutrinos.
The dot-dashed curve shows the result for Dirac NH neutrinos
with the mean square of the relative spin-momentum angle, $\langle\theta^2\rangle$,
given by the estimate for the Milky Way with neutrino magnetic moment $\mu_\nu =  5\times 10^{-14} \mu_B$.
}
\label{figV4}
\end{figure}

The ITBD rate for capturing relic neutrinos is
\beq
\Gamma = \int \sum_{i,h} \frac{d\Gamma^h_i(E_\nu)}{dE_\nu} d E_\nu,
\label{Relic3}
\eeq
summed over the mass state $i$ and helicity states $h$ for both neutrinos and
antineutrinos, and integrated over the energy distribution of the relic
neutrinos. The $i$ and $h$ dependence of $\Gamma^h_i(E_\nu)$ 
in Eq. (\ref{Relic1}) comes from $|U_{ei}|^2 A^h_i$, cf. Eqs.~\eqref{Cr1} and \eqref{Relic1}; we define
\beq
\bar A = \sum_{i,h} |U_{ei}|^2 A^h_i.
\label{Relic4}
\eeq

  For Majorana neutrinos, both neutrinos and antineutrinos can participate
in the ITBD reaction.  With a helicity rotation angle $\theta$, the probablity
of finding a neutrino in a negative or positive helicity state is
$\cos^2(\theta/2)$ and $\sin^2(\theta/2)$, respectively. The corresponding probabilities  for antineutrinos
are $\sin^2(\theta/2)$ and $\cos^2(\theta/2)$. From
Eqs.~(\ref{Cr2}) and (\ref{Relic4}), $\bar A$ for Majorana neutrinos becomes
\beq
\bar A_M & = & \sum_i |U_{ei}|^2 [\cos^2(\theta_i/2) (1+ \beta_i)
+ \sin^2(\theta_i/2) (1- \beta_i) \nonumber \\
 & + & \sin^2(\theta_i/2) (1+ \beta_i)
+ \cos^2(\theta_i/2) (1- \beta_i)] \nonumber \\
 & = & 2 \sum_i |U_{ei}|^2 = 2,
\label{Relic5}
\eeq
where we use the unitarity relation $\sum_i |U_{ei}|^2 = 1$.  
The second line is the antineutrino contribution.
Hence, the ITBD rate for Majorana relic neutrinos is independent of the
mass and helicity of neutrinos.   Dirac antineutrinos, on the other hand,
cannot participate in the ITBD reaction. Hence $\bar A$ can only
come from neutrinos and is
\beq
\bar A_D & = & 1+ \sum_i |U_{ei}|^2 \beta_i \cos\theta_i.
\label{Relic6}
\eeq

From the expressions of the helicity rotation angle $\theta$ given in
Ref.~\cite{Baym1,Baym2}, one can evaluate the ITBD rate for relic neutrinos
on a 100 g tritium target using Eq. (\ref{Relic3}).
Figure~\ref{figV4} shows the ITBD rate for relic neutrinos 
as a function of the mass of the lightest neutrino,
for both Dirac and Majorana neutrinos
with normal and inverted mass hierarchies.
For Majorana neutrinos, the ITBD rate is 8.2/year independent of the
mass hierarchy and the mass of the lightest neutrino. For 
Dirac neutrinos maintaining their original helicity
($\theta = 0$), the ITBD rates are shown as the black and red
curves, respectively, in Fig.~\ref{figV4}. 
As the mass of the lightest neutrino approaches
zero, $\bar A_D$ approaches $1 + |U_{e1}|^2 = 1.6794$
in the normal and $1 + |U_{e3}|^2 = 1.0216$ in the
inverted hierarchy. When the lightest neutrino mass exceeds
0.1 eV, all neutrinos become nonrelativistic and 
$\bar A_D$ approaches unity regardless of the mass hierarchy.  Since $\bar A_D < \bar A_M$, the ITBD rate for Dirac neutrinos is
always smaller than that for Majorana neutrinos.

The dashed curves in Fig.~\ref{figV4} also show the dependence of the
ITBD rate on the lightest (Dirac) neutrino mass for complete helicity flip,
$\theta = \pi$.  For partial spin rotation, the ITBD rate lies
between the solid and dashed curves, while for complete randomization 
of the neutrino helicity, the rate is $\sim$ 4.1/year, just half that for 
Majorana neutrinos. 
 The dot-dashed curve in Fig.~\ref{figV4}
corresponds to an ITBD rate for Dirac relic neutrinos passing through
the Milky Way with a magnetic moment
$\mu_\nu = 5\times10^{-14} \mu_B$, almost three orders of magnitude
smaller than the magnetic moment XENON1T suggest as a possible 
explanation of their low energy
event excess. If the magnetic moment
of normal hierarchy Dirac neutrinos is of the order suggested
by XENON1T,  then 
the neutrino spin rotations would be very large with a broad distribution
and the ITBD rate would approach $\sim$ 4.1/year.

In summary, we have examined the prospect of using an intense $^{51}$Cr
source for observing the inverse tritium beta decay for the first time.
A 3-MCi $^{51}$Cr source placed at a distance of 50 cm from a 100 g
tritium target, would give an expected rate $\sim 150$ per year, more 
than a factor of 10 greater than the expected ITBD rate for relic neutrinos. 
The electrons from
the ITBD with a $^{51}$Cr source will have energies well separated from the
end-point energy of tritium beta decay.  Thus the requirement on the
energy resolution of the spectrometer for separating the $^{51}$Cr ITBD
signals from the tritium beta decay background is greatly relaxed, and
a tritium target much thicker than the PTOLEMY target could be used
for observing $^{51}$Cr ITBD events at a much higher rate.
The PTOLEMY tritium target is designed to have a thickness of
$1 \mu$g/cm$^2$ in order to achieve sub-eV energy resolution for
electron detection.  Conceivably an additional tritium
target with thickness $\sim 0.1$ g/cm$^2$, a factor $10^5$
thicker than the currently designed PTOLEMY target, can be utilized
for a dedicated measurement of the ITBD signals from the $^{51}$Cr
source.  This measurement,  which only requires an energy resolution $\sim 0.1$ MeV,
could lead to the first observation of the ITBD reaction prior to
a search for relic neutrinos.

   We have also investigated
the implications of the relic neutrino helicities on their detection in the
ITBD reaction, and note a significant helicity modification even
if $\mu_\nu$ is two orders of magnitude smaller than the Borexino 
upper limit. We emphasize that the ITBD rate for
relic neutrinos depends on whether the neutrino is Dirac or Majorana,
on the mass hierarchy and the magnetic moment of neutrinos, and on 
the helicity modifications by cosmic gravitational perturbations.
However, it becomes increasingly difficult to resolve the relic
neutrino events from the tritium beta decay background for smaller
neutrino mass, and extremely fine energy resolution of the detector, 
on the sub-eV scale, would be
required. New detector techniques are needed to fully explore the
region of interest shown in Fig. 4.

We also find that the expected ITBD rate from solar neutrinos 
is a factor
of $\sim$ 5-10 smaller than that for relic neutrinos. Thus, unlike the
``neutrino floors" imposed by the solar neutrinos in the search for
dark matter, the search for relic neutrinos via the ITBD reaction is 
not limited by the presence of the solar neutrinos.

This research was support in part by the NSF Grant No. PHY18-12377.

\end{document}